\documentclass[11pt]{article}
\usepackage{amsmath,amssymb,color}

\usepackage{amsfonts}

\newcommand{\hoch}[1]{$\, ^{#1}$}

\makeatletter
\@addtoreset{equation}{section}
\makeatother

\newcommand{\be}{\begin{equation}}
\newcommand{\ee}{\end{equation}}
\newcommand{\bea}{\setlength\arraycolsep{2pt} \begin{eqnarray}}
\newcommand{\eea}{\end{eqnarray}}
\newcommand{\nn}{\nonumber}

\def\ft#1#2{{\textstyle{\frac{\scriptstyle #1}{\scriptstyle #2} } }}
\def\fft#1#2{{\frac{#1}{#2}}}

\def\0{{\sst{(0)}}}
\def\1{{\sst{(1)}}}
\def\2{{\sst{(2)}}}
\def\3{{\sst{(3)}}}
\def\4{{\sst{(4)}}}
\def\5{{\sst{(5)}}}
\def\6{{\sst{(6)}}}
\def\7{{\sst{(7)}}}
\def\8{{\sst{(8)}}}
\def\sst#1{{\scriptscriptstyle #1}}

\def\del{{\partial}}
\def\R{{\mathbb{R}}}

\def\cH{{{\cal H}}}
\def\cQ{{{\cal Q}}}

\begin{document}

\begin{flushright}
\hfill{ \
MIFPA-13-24\ \ \ \ }
\end{flushright}

\vspace{25pt}
\begin{center}
{\Large {\bf AdS Dyonic Black Hole and its Thermodynamics}
}

\vspace{30pt}

{\Large
H. L\"u\hoch{1}, Yi Pang\hoch{2} and C.N. Pope\hoch{2,3}
}

\vspace{10pt}

\hoch{1}{\it Department of Physics, Beijing Normal University,
Beijing 100875, China}

\vspace{10pt}

\hoch{2} {\it George P. \& Cynthia Woods Mitchell  Institute
for Fundamental Physics and Astronomy,\\
Texas A\&M University, College Station, TX 77843, USA}

\vspace{10pt}

\hoch{3}{\it DAMTP, Centre for Mathematical Sciences,
 Cambridge University,\\  Wilberforce Road, Cambridge CB3 OWA, UK}

\vspace{20pt}

\underline{ABSTRACT}
\end{center}
\vspace{15pt}

We obtain spherically-symmetric and $\R^2$-symmetric dyonic black holes that
are asymptotic to anti-de Sitter space-time (AdS), which are solutions
in maximal gauged four-dimensional supergravity, with just
one of the $U(1)$ fields carrying both the
electric and magnetic charges $(Q,P)$.  We study the thermodynamics, and
find that the usually-expected first law does not hold unless $P=0$,
$Q=0$ or $P=Q$.
For general values of the charges, we find that the first law requires
a modification with a new pair of thermodynamic conjugate variables.
We show that they describe the scalar hair that breaks some of the asymptotic AdS symmetries.

\thispagestyle{empty}

\pagebreak
\voffset=-40pt
\setcounter{page}{1}

\tableofcontents

\addtocontents{toc}{\protect\setcounter{tocdepth}{2}}



\section{Introduction}

The spherically-symmetric dyonic black hole of the four-dimensional
Einstein-Maxwell-Dilaton theory obtained by the Kaluza-Klein reduction of
five-dimensional pure gravity has some
intriguing properties \cite{dyon1,dyon2,dyon3,dyon4,dyon5,dyon6}.  The
solution can be obtained by means of a
solution-generating technique, by performing a further $S^1$ reduction of the
four-dimensional theory, which yields a scalar $SL(3,\R)/O(3)$ sigma model
in three dimensions. Acting on the reduction of an uncharged black hole
with an appropriate $O(1,1)\times O(1,1)\in SL(3,\R)$ transformation
then yields the dyonic black hole after lifting back to
four dimensions.  One can alternatively construct the solution by solving
the equations of motion directly; they can be reduced to a set of
$SL(3,\R)$ Toda equations, which can be solved exactly \cite{toda}.
The generalization to arbitrary dimensions and for an arbitrary $SL(n,\R)$
Toda system were given recently in \cite{LY}.  In the extremal limit
the near-horizon geometry is AdS$_2\times S^2$, and the mass is related 
to the charges by the somewhat unusual formula
\begin{equation}
M=(P^{2/3} + Q^{2/3})^{3/2}\,.
\end{equation}
Thus the black hole describes a bound state of electric and magnetic
monopoles with negative binding energy. Although the solution is not
supersymmetric, the entropy $S=8\pi P Q$ in the extremal limit can be
derived from the attractor mechanism \cite{Goldstein:2005hq},
which is analogous to the procedure for supersymmetric solutions.  In fact, the
entropy of a general extremal charged black hole in the
STU model is given by the triality-invariant expression of
the form \cite{kalkol,cvhu} $S=8\pi \sqrt{|D(p,q)|}$, where
\begin{equation}
D(p,q)= 4[(p^1 q_1)(p^2 q_2) + (p^1 q_1)(p^3 q_3) +
  (p^2 q_2)(p^3 q_3) -p^0 q_1 q_2 q_3 +q_0 p^1 p^2 p^3] -(p^\mu q_\mu)^2\,,
\label{duffform}
\end{equation}
where $q_\mu$ are the electric charges and $p^\mu$ the magnetic charges for
the four gauge fields.  The last term gives the contribution for the
Kaluza-Klein dyon.  The solution can be lifted to become a D0/D6 brane
in string theory,
or a pp-wave/NUT intersection in M-theory.

   In this paper, we consider the Kaluza-Klein theory with an added
scalar potential.
To be more precise, the theory we consider can be embedded into maximal
gauged supergravity in four
dimensions.
Maximal gauged supergravity in four dimensions has an $SO(8)$
gauge group, which has a $U(1)^4$ Cartan subgroup.  In this paper,
we set all except one of these $U(1)$ gauge fields to zero.  The minimal
bosonic field content then involves the metric, a dilaton $\phi$ and the vector
field $A_\mu$.  The relevant Lagrangian is given by
\begin{equation}
{\cal L}=\sqrt{-g}\Big[R - \ft12 (\partial\phi)^2 - \ft14 e^{-\sqrt3\,\phi} F^2 +6g^2 \cosh\Big(\ft1{\sqrt3} \phi\Big)\Big]\,,\label{lag}
\end{equation}
where $F=dA$.  (There should be no confusion between the gauge coupling
constant $g$ and the determinant of the metric.)  The scalar potential has
a stationary point at $\phi=0$, giving rise to a supersymmetric
four-dimensional anti-de Sitter (AdS$_4$) vacuum, with AdS
radius $\ell=1/g$. If we set the gauge coupling constant $g$ to zero, the
theory is the standard Kaluza-Klein theory obtained from the $S^1$ reduction
of pure gravity in five dimensions.  For non-vanishing $g$, the theory
can be obtained as a consistent truncation of the $S^7$ reduction of
eleven-dimensional supergravity.

    Although the theory (\ref{lag}) cannot be supersymmetrised on
its own, it
can be ``pseudo-supersymmetrised.''  In other words, by adding a
pseudo-gravitino and pseudo-dilatino, the Lagrangian can be made
invariant, up to quadratic order in fermions, under pseudo-supersymmetry
transformations.  This conclusion can be generalized to the
Kaluza-Klein theory in an arbitrary dimension, for a specific choice
of the scalar potential \cite{Liu:2012jra,Liu:2011ve}.  In $D=4,5,6$ and 7
dimensions, the pseudo-supersymmetry transformation rules coincide with
those in known supergravities in those dimensions, since the theories can
then all be embedded in the relevant supergravities.

In section 2, we present the local spherically-symmetric solution.
It is asymptotically AdS, and it contains three non-trivial parameters,
characterising the mass $M$, and the electric and magnetic charges $Q$ and
$P$. It should be emphasised that there is no continuous duality
rotation symmetry in the theory, and so the generic dyonic black hole
cannot be simply rotated into a purely electric or magnetic one.

   By looking at the asymptotic form of the metrics, we obtain
expressions for the conserved quantities $M$, $Q$ and $P$ in terms of the
original parameters in the local metric.  We then investigate the
thermodynamics, and we find that the naive form $dM=TdS + \Phi_Q \,dQ +
\Phi_P\, dP$ of the first law
is not obeyed unless one or another of the following conditions holds:
\begin{eqnarray}
\hbox{Case I}:&& g=0\,,\qquad \hbox{for general $(Q,P)$}\,;\cr
\hbox{Case II}:&& g\ne 0\,,\qquad \hbox{and}\qquad Q=0\hbox{\ \ or\ \ }
P=0 \hbox{\ \  or\ \ }Q=P\,.
\label{exception}
\end{eqnarray}
For the general AdS dyons, we find that it is necessary to introduce a
new thermodynamic quantity $Y$ and its conjugate $X$, so that
\be
dM= TdS + \Phi_Q\, dQ + \Phi_P\, dP + X dY\,,
\ee
in order to obtain a first law.  We also verify that the product of
the ``entropies'' at all of the horizons obeys a relation that is in
accordance with general results seen in other black hole example, depending
only on the quantised charges $Q$ and $P$ and the gauge coupling $g$.

  In section 3, we take an appropriate limit to obtain the AdS dyonic
black hole with a flat $\R^2$ horizon rather than the original $S^2$.
The asymptotic form of the metric is then AdS in Poincar\'e planar coordinates.
Again we find that the first law of thermodynamics requires the introduction
of the additional conjugate pair $(X,Y)$ of thermodynamic variables.

     In section 4, we consider various limits of the AdS dyons.
The extremal limit gives
rise to near-horizon geometries that are either AdS$_2\times S^2$ or
AdS$_2\times \R^2$.   For the spherically-symmetric solution, the
``BPS''-like limit, which would give an extremal black hole in the case
that $g=0$,
now gives rise to a solution with a naked singularity.  For the
$\R^2$-symmetric solution, the gauge potential vanishes in the extremal limit,
giving rise to a domain wall (membrane) solution supported by the
potential for the scalar field.  In section 5, we study the lift of the
solutions to $D=11$, via the consistent $S^7$ reduction.

    In section 6, we investigate the origin of the thermodynamic pair
$(X,Y)$ that we had to introduce in the first law.
By studying the asymptotics, we find that
the fall-off of the scalar fields breaks some of the asymptotic AdS
symmetries,  except in the special cases enumerated in (\ref{exception})
for which the extra variables $X$ and $Y$ are not needed.   We then
employ Wald's procedure and obtain a scalar potential and charge,
which are precisely the quantities $(X,Y)$ that are needed for
establishing the first law of thermodynamics.  We conclude the paper with
comments in section 7.

\section{$S^2$-symmetric Dyonic AdS Black Hole}

In this section we obtain the spherically-symmetric dyonic black hole solution
of the theory described by (\ref{lag}), carrying both electric and
magnetic charges.  We shall first present the local solution and then
discuss its thermodynamics.

\subsection{Local solutions and conserved quantities}

We find that the Lagrangian (\ref{lag}) admits the following static solution
\begin{eqnarray}
ds^2 &=& -(H_1 H_2)^{-\fft12} f dt^2 + (H_1 H_2)^{\fft12} \Big(\fft{dr^2}{f} + r^2 (d\theta^2 + \sin^2\theta\, d\varphi^2)\Big)\,,\cr
\phi&=&\fft{\sqrt{3}}{2} \log\fft{H_2}{H_1}\,,\qquad
f=f_0 + g^2 r^2 H_1 H_2\,,\qquad f_0=1 - \fft{2\mu}{r}\,,\cr
A&=&\sqrt2 \Big(\fft{(1 - \beta_1 f_0)}{\sqrt{\beta_1\gamma_2}\, H_1}\, dt + 2\mu\,\gamma_2^{-1}\sqrt{\beta_2\gamma_1}\, \cos\theta\, d\varphi\Big)\,,\cr
H_1&=&\gamma_1^{-1} (1-2\beta_1 f_0 + \beta_1\beta_2 f_0^2)\,,\qquad
H_2=\gamma_2^{-1}(1 - 2\beta_2 f_0 + \beta_1\beta_2 f_0^2)\,,\cr
\gamma_1&=& 1- 2\beta_1 + \beta_1\beta_2\,,\qquad \gamma_2 = 1-2\beta_2 + \beta_1\beta_2\,.
\label{adsdyon}
\end{eqnarray}
The solution is written in a form where the parameters
$\beta_i$ that characterise the charges enter in a symmetrical
way, as was done in \cite{LY} for the $g=0$ (ungauged) case.  The charge
is purely magnetic if we set
$\beta_1=0$, and purely electric if instead we set $\beta_2=0$.
If we set $\beta_1=\beta_2$, the dilaton decouples and the metric becomes
a dyonic Reissner-Nordstr\"om AdS black hole.

   For the general case,
one must require that $\beta_i>0$ and $\gamma_i\ge 0$ for it to be well
behaved from the horizon out to infinity.  This implies that one cannot
freely choose the ranges of $\beta_1$ and $\beta_2$ independently.  A different
parameterisation, in terms of two independently-specifiable quantities
$\lambda_1$ and $\lambda_2$, is given by writing
\begin{equation}
\beta_i=\lambda_j^{-1} (1 - \sqrt{1-\lambda_1 \lambda_2})\,,
\qquad i \ne j\,,\label{lambdas}
\end{equation}
The parameters $\lambda_i$ can then independently take values lying in
the range $0\le\lambda_i\le 1$.

      Our construction of this solution proceeded as follows.
We first considered the case with vanishing gauge coupling constant, $g=0$.
This Kaluza-Klein dyon has long been known.  It turns out that the
equations of motion for static black holes
can be reduced to a system of one-dimensional
$SL(3,\R)$ Toda equations, which are completely solvable.
(See, for example, \cite{toda,LY}.)  The general solution contains
four integration constants.  One of them is associated with
a non-trivial scalar deformation and this has to be removed since it either
gives rise to a naked curvature singularity or else it destroys the
asymptotic flatness of the space-time.   We adopted the notation of
\cite{LY} in presenting the solution, and the parameters
$(\mu, \beta_1,\beta_2)$ can be thought of as characterising the mass
and the electric and magnetic charges.  An alternative approach
for constructing the ungauged dyonic black hole is to use a
solution-generating technique.  For $g=0$, the $S^1$ reduction of the
Lagrangian (\ref{lag}), after dualising the Kaluza-Klein vector to an axion,
gives a non-linear sigma model for the coset $SL(3,\R)/SO(3)$.  Starting from
an uncharged four-dimensional black hole, and reducing it to three dimensions,
one then uses the two $O(1,1)$ boosts in $SL(3,\R)$ to generate a new
solution which, after lifting back to four dimensions, describes the
dyonic black hole with electric and
magnetic  charges.  This technique can be used to construct rotating
dyonic black holes, as first obtained in \cite{dyon6}, but for our
purposes we need only the static case.
The two $O(1,1)$ boost parameters $\delta_i$ are related to the
$\beta_i$ we are using here by\footnote{In the solution-generating process it
is necessary also to make a compensating $U(1)$ transformation to remove
an unwanted NUT charge, and this results eventually in a somewhat awkward
parameterisation in terms of the boost parameters $\delta_i$, which is why
we prefer to use the $\beta_i$ parameterisation here.}
\begin{equation}
\sinh^2\delta_1 = \fft{2\beta_1 \gamma_2 }{(1+\beta_1\beta_2)\gamma_1}\,,\qquad
\sinh^2\delta_2= \fft{2\beta_2}{\gamma_2}\,,
\end{equation}

   There is no solution-generating technique available for constructing the
solution in the gauged theory.  However, we found that
having obtained the solution with $g=0$, we can simply replace the function
$f_0$ by $f$ in the metric in order to obtain the solution in
the $g\ne0$ theory.  This construction is analogous to that in other
charged static AdS black holes in gauged supergravities
\cite{Behrndt:1998jd,Duff:1999gh,tenauthor,Cvetic:1999un} and in
certain general classes of non-supersymmetric theory \cite{Lu:2013eoa}.

    The solution is asymptotic to AdS written in global coordinates.
The mass can be easily calculated using the
AMD method developed in \cite{ashmag,ashdas},
by studying the asymptotic fall-off of the Weyl tensor on the conformal
boundary (see also \cite{Chen:2005zj} for a discussion of the AMD procedure
for  asymptotically AdS black holes in supergravities). We find that it
 is given by
\begin{equation}
M= \fft{(1-\beta_1)(1-\beta_2)(1-\beta_1\beta_2) \mu}{\gamma_1 \gamma_2}\,.
\label{mass}
\end{equation}
The electric and magnetic charges $Q$ and $P$ can be obtained
straightforwardly by integrating the conserved quantities, giving
\begin{equation}
Q = \fft{\mu\,\sqrt{\beta_1\,\gamma_2}}{\sqrt2\, \gamma_1}\,,\qquad
P= \fft{\mu\, \sqrt{\beta_2\, \gamma_1}}{\sqrt2\, \gamma_2}\,.\label{QPdef}
\end{equation}
Note that these conserved charges are independent of the gauge
coupling constant $g$ and are the same as those given in \cite{LY}
for the case $g=0$.

   For later purpose, it is instructive to study the asymptotic fall-off.
To do this, we introduce a new radial coordinate $\rho$, such that the
metric becomes
\begin{equation}
ds^2=-h dt^2 + \tilde h^{-1}\, d\rho^2 + \rho^2 d\Omega_2^2\,.\label{schcoord}
\end{equation}
For large $\rho$, we find
\begin{eqnarray}
h&=& g^2 \rho^2 + 1 - \fft{2M}{\rho} + \fft{4(Q^2 + P^2)}{\rho^2}
  + \cdots\,,\cr
\tilde h &=& g^2\rho^2 + 1 +
 \fft{3g^2\mu^2(\beta_1-\beta_2)^2(1-\beta_1\beta_2)^2}{\gamma_1^2\gamma_2^2}\cr
&& - \fft{1-\beta_1\beta_2}{\gamma_1\gamma_2} \Big[2\mu(1-\beta_1)(1-\beta_2)
+\fft{8g^2\mu^3(\beta_1-\beta_2)^2 \gamma}{\gamma_1^2\gamma_2^2}\Big]\fft{1}{\rho} + \cdots\,,\cr
 &=& h\Big[1 + \fft{3\mu^2(\beta_1-\beta_2)^2(1-\beta_1\beta_2)^2}{\gamma_1^2
\gamma_2^2\,\rho^2} - \fft{8\mu^3 (\beta_1-\beta_2)^2(1-\beta_1\beta_2)
\gamma}{\gamma_1^3\gamma_2^3\,\rho^3} +\cdots\Big]\,,
\end{eqnarray}
where the constant $\gamma$ is defined by
\begin{equation}
\gamma\equiv\beta_1 + \beta_2 - 8 \beta_1 \beta_2+6 \beta_1^2 \beta_2 +
6 \beta_1 \beta_2^2 - 8 \beta_1^2 \beta_2^2 + \beta_1^3 \beta_2^2 +
\beta_1^2 \beta_2^3\,.
\end{equation}
The scalar at large $\rho$ is given by
\begin{equation}
\cosh\Big(\fft\phi{\sqrt3}\Big)=1 + \fft{2\mu^2(\beta_1-\beta_2)^2(1-\beta_1\beta_2)^2}{\gamma_1^2\gamma_2^2\,\rho^2} -
\fft{4\mu^3(\beta_1-\beta_2)^2(1-\beta_1\beta_2) \gamma}{\gamma_1^3\gamma_2^3\rho^3} + \cdots\,.
\end{equation}
The scalar $\phi$ vanishes if $\beta_1=\beta_2$, and consequently we
have $h=\tilde h$ and the metric becomes that of the Reissner-Nordstr\"om
black hole.  (The scalar does not vanish for $\beta_1\beta_2=1$,
which corresponds to a naked singularity.)

\subsection{Thermodynamics}

In the previous subsection, we calculated the mass and the electric and
magnetic charges of the gauged dyonic solution.  These quantities were
determined from the asymptotic form of the solution near infinity,
irrespective of
whether the solution had a regular black hole horizon or not.  For the
solution to describe a black hole, the parameters must be restricted so that
the metric describes an event horizon at some radius $r=r_0$.   All the
fields, including the metric, gauge field and scalar, should be well
behaved from $r=r_0$ to $r\rightarrow \infty$.  The horizon is defined
to be the largest (real) root of the function $f$, namely
\begin{equation}
1-\fft{2\mu}{r_0} + g^2 r_0^2 H_1(r_0) H_2(r_0)=0\,.
\end{equation}
Note that for $g=0$, the outer horizon is simply located at $r_0=2\mu$,
in which case
the inner horizon is at $r=0$.  For non-vanishing $g$, none of the
the quantities $\beta_i$, $\mu$ or $r_0$ can conveniently
be solved for in closed form.  Furthermore, the surface at $r=0$ plays
no distinguished role, and it does
not correspond to a null surface.

   The temperature and the entropy associated with the outer horizon can be
obtained using standard techniques, and are given by
\begin{equation}
T= \fft{f'(r_0)}{4\pi \sqrt{H_1(r_0) H_2(r_0)}}\,,\qquad S=\pi r_0^2
\sqrt{H_1(r_0)H_2(r_0)}\,.\label{tempentro}
\end{equation}
The calculation of the electric potential $\Phi_Q$ is straightforward; it is
given by the
time component of the potential $A_\mu dx^\mu$. The magnetic
potential $\Phi_P$
can be obtained by performing a Hodge dualisation, which exchanges
the roles of the electric and magnetic charges.  The magnetic potential is
then given by the time component of the vector potential for the dual field.
Thus we find that the electric and magnetic potentials, which vanish at
infinity, are given on the horizon by
\begin{eqnarray}
\Phi_Q &=& \sqrt{\fft{2}{\beta_1\gamma_2}}
\Big(1 - \beta_1 - \fft{1-\beta_1 f_0(r_0)}{H_1(r_0)}\Big)\,,\cr
\Phi_P &=& \sqrt{\fft{2}{\beta_2\gamma_1}}
\Big(1 - \beta_2 - \fft{1-\beta_2 f_0(r_0)}{H_2(r_0)}\Big)\,.
\end{eqnarray}

  One might expect that the first law of thermodynamics should be just
$dM=T dS + \Phi_Q \,dQ + \Phi_P \, dP$, but, as mentioned in the introduction,
this is not in general the case, with the exceptions listed in
(\ref{exception}).  In particular, it does hold for generic
electric and magnetic charges provided that the gauge-coupling $g$ is zero.
This is to be expected, since the thermodynamics of the asymptotically flat
Kaluza-Klein dyon has long been understood.  (See, for example, \cite{dyon6}.)
For non-vanishing $g$, we find the first law holds only if $\beta_1=0$,
corresponding to $Q=0$; or if $\beta_2=0$, corresponding to $P=0$;
or if
$\beta_1=\beta_2$, corresponding to $Q=P$.  In order for the first law
to work for
generic choices of the charges, we must introduce an
additional intensive/extensive conjugate pair of thermodynamic
variables $(X,Y)$, and modify the first law to become
\begin{equation}
dM=T dS + \Phi_Q\, dQ + \Phi_P\, dP + X dY\,.\label{modfirst}
\end{equation}
By solving for $X$ and $Y$ such that this modified law holds, we find that
$X$ and $Y$ can be taken to be
\begin{equation}
X=\fft{4g^2\mu^3(\beta_1-\beta_2)\sqrt{\beta_1\beta_2^3}}{(1-\beta_1\beta_2)
\gamma_2^2} \,,\qquad
Y=\fft{\sqrt{\beta_1}\, \gamma_2}{\sqrt{\beta_2}\, \gamma_1}\,,
\end{equation}
There is actually a certain freedom in our choice of the functions
$X$ and $Y$.  We can, accordingly, replace $Y$ by any function of $Y$,
provided we then make the corresponding appropriate adjustment to $X$.
Although $Y$ becomes singular when $\beta_2=0$, the quantity $XdY$
vanishes and hence the extra ``hair'' does not contribute to the
thermodynamics in this case.  In fact, it is easy to see that $XdY$
has a zero contribution if either $g=0$, or else, when $g\ne 0$, if
$\beta_1=0$ or $\beta_2=0$ or $\beta_1=\beta_2$.  In other words,
$XdY$ plays no role in any of the cases enumerated in (\ref{exception})
for which the usual first law holds.  We shall give a detailed discussion
of the significance of the $X$ and $Y$ variables in section 6.

   We can also go one step further, and treat the cosmological constant
$\Lambda=-3g^2$ of the asymptotically-AdS metric
as an additional thermodynamic quantity like a pressure, with
its conjugate ``potential'' $\Upsilon$, such that
\begin{equation}
dM=T dS + \Phi_Q\, dQ + \Phi_P\, dP + X dY + 
\Upsilon d\Lambda\,.\label{firstlaw}
\end{equation}
We find that $\Upsilon$, which can be interpreted as a volume
(see, for example, \cite{cvgikupo} and references therein), is given by
\begin{eqnarray}
\Upsilon&=&\fft{r_0^3}{12\gamma_1\gamma_2} \Big(\beta_1 \beta_2 (-\beta_1 -
\beta_2 + 2 \beta_1 \beta_2) \tilde f_0^3
-3 \beta_1 \beta_2 (2 - \beta_1 - \beta_2) \tilde f_0^2\cr
 &&\qquad\qquad+ 3 (\beta_1  \beta_2 - 2 \beta_1 \beta_2) \tilde f_0 -2 +
 \beta_1 + \beta_2\Big)\,,
\end{eqnarray}
where $\tilde f_0=1-\mu/r_0$.  

        If we define a thermodynamic potential
\begin{equation}
\Phi_{\rm thermo} \equiv M - T S - \Phi_Q \, Q - \Phi_P\, P\,,
\label{thermopot1}
\end{equation}
then we find it is given simply by
\be
\Phi_{\rm thermo}= \ft12(r_0-\mu)\,.\label{thermopot2}
\ee
It does not explicitly depend on $g$ and $\beta_i$, a feature
that appears to be a
rather universal for the general class of $SL(n,\R)$ black holes
considered in \cite{LY}.  We also find that the Smarr formula is given by
\begin{equation}
M=2 T S + \Phi_Q\,  Q + \Phi_P\,  P - 2 \Upsilon \Lambda\,.\label{smarr}
\end{equation}
Note that, given the first law (\ref{firstlaw}), the coefficient of the terms
in (\ref{smarr}) follow just from the scaling dimensions of the
thermodynamic quantities, and in particular, since $Y$ is dimensionless,
$XY$ does not appear in the Smarr formula.

\subsection{Euclidean action and free energy}

    We may compute the Euclidean action by using the holographic 
renormalization method described in \cite{de Haro:2000xn}. The divergences 
in the classical on-shell Euclidean action are removed by the addition of
local covariant surface counterterms. In our case, the renormalized 
action in Lorentzian signature is given by
\be
I=I_{\rm bulk}+I_{\rm surf}+I_{\rm ct}\,,\label{Iren}
\ee
where 
\bea
I_{\rm bulk} &=& \frac1{16\pi G}\int_{\cal M} d^4x\sqrt{-g}
\Big(R - \ft12 (\partial\phi)^2 - \ft14 e^{-\sqrt3\,\phi} F^2 +
   6g^2 \cosh(\ft1{\sqrt3} \phi)\Big)\,,\nn\\
I_{\rm surf} &=&-
 \frac{1}{8\pi G}\int_{\partial {\cal M}}d^3x \sqrt{-h}K\,,\label{action}
\eea
and the counterterms take the form
\be
I_{\rm ct}=-\frac{1}{8\pi G }\int_{\partial {\cal M}} d^3x
 \sqrt{-h}\Big(\frac2{\ell}+\frac{\ell}2{\cal R}\Big)
+\frac{1}{48\pi G} \int_{\partial {\cal M}} d^3x \sqrt{-h}
\Big(\phi n^{\mu}\partial_{\mu}\phi-\frac1{2\ell}\phi^2
\Big)\,.
\ee
In the above equation, $K_{\mu\nu}\equiv
  -\ft12 (\nabla_{\mu}n_{\nu}+\nabla_{\nu}n_{\mu})$ is the extrinsic 
curvature of the boundary surface, with $n_{\mu}$ being the outward 
unit normal vector. $\ell=1/g$ is the curvature radius of AdS, and
${\cal R}$ is the Ricci scalar of the boundary metric.

  Using these counterterms, the renormalized energy-momentum tensor 
is given by $T^{\mu\nu}\equiv (2/\sqrt{-h})\, \delta I/\delta h_{\mu\nu}$,
yielding
\be
T_{\mu\nu}=\frac1{8\pi G}
\Big(K_{\mu\nu}-Kh_{\mu\nu}-\frac2{\ell}h_{\mu\nu}+
\ell \, ({\cal R}_{\mu\nu} -\ft12 {\cal R}\, h_{\mu\nu})
+\frac16 h_{\mu\nu}(\phi n^{\rho}\partial_{\rho}\phi-\frac1{2\ell}\phi^2)
\Big)\,.
\ee
In cases where the boundary geometry possesses a timelike Killing vector, 
one can define the energy of the dual field theory, which is equal to the 
energy of the black hole via the AdS/CFT correspondence. 

  We parameterise 
the boundary metric as a foliation of spacelike surfaces $\Sigma$ with 
metric $\sigma_{ab}$:
\be
h_{\mu\nu}dx^{\mu}d^{\nu}=-N^2dt^2+\sigma_{ab}(dx^a+N^adt)(dx^b+N^bdt)\,.
\ee
The energy can be calculated by evaluating
\be
M=\int_\Sigma d^2x\sqrt{\sigma}Nu^{\mu}u^{\nu}T_{\mu\nu}\,,
\ee
where $u^{\mu}$ is the the unit timelike normal to $\Sigma$.  Evaluating 
this for the AdS dyon (\ref{adsdyon}), we recover precisely the AMD
expression for the mass that we obtained in (\ref{mass}).  

   By definition the Euclidean action $I_E$ is given by the
Euclideanisation of $-I$ defined in (\ref{Iren}), and by standard arguments 
\cite{bromaryor} this is equal to $\beta\, (\cH|_\infty - \cH|_{r_0})$,
where $\cH|_\infty$ and $\cH|_{r_0}$ are the values of the Hamiltonian 
at infinity and at the horizon $r=r_0$.  Since our electrostatic potential is
chosen to vanish at infinity we then find
\be
\cH|_\infty = M\,,\qquad \cH|_{r_0} = T S + \Phi_Q\, Q\,,
\ee
and hence we arrive at the result that
\be
T I_E=M-TS-\Phi_Q \, Q\,.\label{free0}
\ee
We see, in particular, that the thermodynamic quantities $X$ and $Y$
do not enter in this relation.  

   It should be noted also that while
the electric charge and electric potential enter in the free energy on the
right-hand side of (\ref{free0}), the magnetic charge and magnetic
potential do not.  We obtained this result using the conventional
definition of the renormalised action (\ref{Iren}).  One could always
add an additional surface term, 
\be
I_{\rm extra}=\fft1{16\pi G} \int_{\del M}
\sqrt{-h} d^3x\, e^{-\sqrt3\, \phi}\, n_\mu F^{\mu\nu}A_\nu\,,
\ee
which would correspond to changing the boundary conditions in the
variational problem from the case where $A_\mu$ is specified on
$\del M$ to the case where instead $F_{\mu\nu}$ is specified.  This
would have the effect of removing the $\Phi_Q\, Q$ term in (\ref{free0}).
(See \cite{chemjomy} for a related discussion.)
This
amounts to performing a Legendre transformation on the free
energy that appears on the right-hand side of (\ref{free0}), to a new
energy for which $Q$, rather than $\Phi_Q$, is held fixed in the first law. 
Note that the thermodynamic potential defined in eqn (\ref{thermopot1})
corresponds to a different Legendre transformation of the free energy in 
(\ref{free0}).  This one, however, could not be achieved by adding a
surface term to the renormalised action (\ref{Iren}) that was a local
function of the potential $A_\mu$.     

\subsection{Entropy product formula}

When $g=0$, there are two null surfaces, located at $r_+=2\mu$ and $r_-=0$.
If we introduce the  concept of an entropy for each null surface, or horizon,
as one quarter of the area of the horizon, one can see that
\cite{cvlupoprod}
\begin{equation}
S_+ S_- = \ft14 \pi^2P^2 Q^2\,,
\end{equation}
which is quantised and independent of the mass parameter.

Analogous entropy product formulae exist for all the $SL(n,\R)$ Toda
black holes \cite{LY}.  The first observation of such a property was given
in \cite{CYII}.  When $g\ne0$ the metric function $f$ has in general
four roots, each corresponding to a null surface.  It was argued that a
universal entropy product formula should still hold if one considers
the product of the entropies of all null surfaces \cite{cvgipo}.  Indeed,
we find that the result in the present case is independent of the mass:
\begin{equation}
\prod_{i=1}^4 S_i = \fft{64\pi^4}{g^4}\, P^2 Q^2\,.\label{entropyprod}
\end{equation}
This result can be obtained by noting from (\ref{adsdyon}) that
\be
S^2_i = \pi^2 (r^4 H_1 H_2)\Big|_{r=r_i}= \fft{\pi^2}{g^2}\, (2\mu-r_i) r_i\,,
\ee
where $r_i$ is the radius of the $i$'th null surface, which is one of the 
four roots of $f(r_i)=0$.  Thus it follows that $\prod_i S_i^2$ is 
expressible in terms of $\mu$ and the four symmetric multinomials 
\be
\Delta_4=\prod_i r_i\,,\quad \Delta_3= r_1 r_2 r_3+\cdots\,,\quad
\Delta_2 = r_1 r_2 + \cdots\,,\quad \Delta_1 = \sum_i r_i \,.
\ee
Since $f(r_i)=0$ for each $i$ it follows that the $\Delta_i$ are 
expressible purely in terms of the parameters $\beta_1$, $\beta_2$ and $\mu$ on
which $f$ depends, and so $\prod_i S_i$ is necessarily expressible as a 
function only of $\beta_1$, $\beta_2$ and $\mu$.  Comparing with the 
expressions given in (\ref{QPdef}), it then turns out that when
expressed in terms of $Q$ and $P$, the product of entropies is independent
of $\mu$.

\section{$\R^2$-Symmetric Dyonic AdS Black Hole}

  In the previous section we constructed the static spherically-symmetric
black hole gauged dyon.  Since the solutions are asymptotic to AdS, there
will also exist solutions whose level surfaces, and horizon, are
flat $\R^2$, or hyperbolic.  These so called ``topological'' black holes
can easily be obtained from the previous $S^2$ black holes
by making an appropriate scaling of the parameters, namely
\begin{equation}
r\rightarrow \epsilon^{-1/2}\,  r\,,\qquad t\rightarrow
\epsilon^{1/2}\,  t\, \qquad \beta_i \rightarrow \epsilon \, \beta_i\,,
\qquad  \mu \rightarrow \epsilon^{-3/2} \mu\,,
\qquad d\Omega_2^2 \rightarrow \epsilon\,  d\Omega_{2,\epsilon}^2\,.
\end{equation}
Taking $\epsilon=-1$ gives the black hole with hyperbolic horizon.
The limit where  $\epsilon=0$ gives the $\R^2$ black hole, which is of
particular interest because the asymptotically AdS region is then written in
Poincar\'e coordinates, and hence is particularly suited for the discussion of the
AdS/CFT correspondence.  The solution is given by
\begin{eqnarray}
ds^2 &=& -(H_1 H_2)^{-\fft12} f dt^2 + (H_1 H_2)^{\fft12} \Big(\fft{dr^2}{f}
+ r^2 (dx^2 + dy^2)\Big)\,,\cr
\phi&=&\ft{\sqrt{3}}{2} \log\fft{H_2}{H_1}\,,\qquad
f=- \fft{2\mu}{r} + g^2 r^2 H_1 H_2\,,\cr
A&=&\sqrt{2\mu} \Big(\fft{(r + 2\beta_1)}{\sqrt{\beta_1}\, H_1\,r}\, dt
+ 2\sqrt{\beta_2}\, xdy\Big)\,,\cr
H_1&=&1 + \fft{4\beta_1}{r} + \fft{4\beta_1\beta_2}{r^2}\,,\qquad
H_2=1 +\fft{4\beta_2}{r} + \fft{4\beta_1\beta_2}{r^2}\,,
\label{adsdyon2}
\end{eqnarray}
In presenting this $\epsilon=0$ solution, we have made a further
redefinition $\beta_i\rightarrow \beta_i/\mu$.  This new parameterisation
is convenient because then $H_1$ and $H_2$ are independent of $\mu$, 
and so one can then solve linearly for
$\mu$ in terms of the horizon radius $r=r_0$, namely
\begin{equation}
\mu=\ft12 g^2 r_0^3 H_1(r_0) H_2(r_0)\,,
\end{equation}
The temperature and the entropy then take the same form as in
(\ref{tempentro}).  Here we are, for convenience, assuming that the
$\R^2$ coordinates $(x,y)$ have been identified to give a 2-torus of
volume $4\pi$.  One can take any other choice for the volume, with the
understanding that the extensive quantities should be scaled by the
relative volume factor. We find that the remaining thermodynamic 
quantities are
\begin{eqnarray}
&&M=\mu\,,\qquad Q=\sqrt{\fft{\mu\beta_1}{2}}\,,\qquad
P=\sqrt{\fft{\mu\beta_2}{2}}\,,\cr
&&\Phi_Q=\fft{2\sqrt{2\mu\beta_1} (r_0 + 2\beta_2)}{r_0^2 H_1(r_0)}\,,\qquad
\Phi_P=\fft{2\sqrt{2\mu\beta_2} (r_0 + 2\beta_1)}{r_0^2 H_2(r_0)}\,,\cr
&&X = 4 g^2 (\beta_1 - \beta_2) \sqrt{\beta_1 \beta_2^3}\,,\qquad
Y=\sqrt{\fft{\beta_1}{\beta_2}}\,,\qquad \Lambda=-3g^2\,,\cr
&&\Upsilon = -\ft16 (4 \beta_1 \beta_2 (\beta_1 + \beta_2) + 12 \beta_1 \beta_2 r_0
+ 3 (\beta_1 + \beta_2) r_0^2 + r_0^3)\,,\label{thermoquan}
\end{eqnarray}
It is now straightforward to verify that the first law of
thermodynamics (\ref{firstlaw}) is again satisfied.  The Smarr formula
takes the same form as (\ref{smarr}) and the thermodynamic potential
is now simply
\begin{equation}
\Phi_{\rm thermo} =-\ft12\mu=-\ft12M\,.
\end{equation}
The entropy product formula is the same as (\ref{entropyprod}).

\section{Limits of The Gauged Dyonic Black Hole}

    Having obtained the general solution, we may consider various limits
of the parameters. The most obvious one to consider is the extremal limit,
for which the function $f$ acquires a double zero at its largest root.
The near-horizon geometry in this case then becomes AdS$_2\times S^2$,
or AdS$_2\times \R^2$.  The procedure for taking this limit
is straightforward and we shall
not present any explicit results.

    Let us instead consider the ``BPS''-type of ``extremal'' limit, in
which we send $\mu$ to zero while keeping $Q$ and $P$ finite.  (We
refer to it as ``BPS'' because this limiting solution becomes supersymmetric
in either of the cases $Q=0$ or $P=0$.)  The limit can be taken by using the
the parameterisation in terms of $\lambda_i$ introduced in (\ref{lambdas}),
and defining
\be
\lambda_i = 1 - \fft{k^2}{b_i^2}\,, \qquad \mu = 2 k b_1\, b_2\,,
\ee
Taking the limit $k\rightarrow 0$ in the spherically-symmetric dyon
(\ref{adsdyon}) we then find $Q=b_1^3$, $P=b_2^3$ and $M=(b_1^2+b_2^2)^{3/2}$,
and
\begin{eqnarray}
H_1 &=& 1 + \fft{4 Q^{2/3} \sqrt{P^{2/3} + Q^{2/3}}}{r}
       + \fft{8 P^{2/3} Q^{4/3}}{r^2}\,,\cr
&&\\ \nn
H_2 &=& 1 + \fft{4 P^{2/3} \sqrt{P^{2/3} + Q^{2/3}}}{r}
   + \fft{8 P^{4/3} Q^{2/3}}{r^2}\,,\cr
&& \\ \nn
f&=&r^2 + g^2 r^4 H_1 H_2\,.
\end{eqnarray}
The vector potential is now given by
\begin{equation}
A=\fft{Q(4r+ 8 P^{2/3} \sqrt{P^{2/3} + Q^{2/3}})}{
r^2 + 4 Q^{2/3} \sqrt{P^{2/3} + Q^{2/3}}\, r +
8 P^{2/3} Q^{4/3}}\, dt + 4P \cos\theta\, d\phi\,.
\end{equation}
If $g=0$, the horizon with the double root is located at $r=0$,
and the near-horizon geometry is AdS$_2\times S^2$.
When $g\ne 0$, the radius $r=0$ has no particular significance, and
the singularity, corresponding to having $H_1 H_2=0$, becomes naked.
The mass and the conjugate pair $(X,Y)$ are given by
\begin{equation}
M=\Big(P^{2/3} + Q^{2/3}\Big)^{3/2}\,,
\qquad X=\fft{g^2 P^2 Q^{2/3} (Q^{2/3} -
       P^{2/3})}{\sqrt{P^{2/3} + Q^{2/3}}}\,,\qquad
       Y=\fft{Q^{2/3}}{P^{2/3}}\,.
\end{equation}

    The situation is very different for the $\R^2$-symmetric AdS dyon
(\ref{adsdyon2}).
In this case, setting $\mu=0$ implies that the gauge potential vanishes,
and the resulting solution becomes the AdS membrane
\begin{eqnarray}
ds^2=r^2(H_1 H_2)^{\fft12}(-g^2 dt^2 + dx^2 + dy^2) + (H_1 H_2)^{-\fft12}
\fft{dr^2}{g^2 r^2}\,,\qquad e^{\fft{2}{\sqrt3}\phi}=\fft{H_2}{H_1}\,,
\end{eqnarray}
where the $H_i$ are again given by (\ref{adsdyon2}).  

\section{Lifting to M-theory}

   The four-dimensional theory described by (\ref{lag}) can itself be
embedded
within the gauged four-dimensional supergravity STU model, whose
bosonic embedding as a 7-sphere reduction of eleven-dimensional
supergravity was given in \cite{tenauthor} (see section 3 of that paper).
The metric reduction ansatz is given by
\be
d\hat s_{11}^2 = \tilde \Delta^{2/3}\, ds_4^2 + g^{-2}\, \tilde\Delta^{-1/3}\,
\sum_{i=1}^4 X_i^{-1} \, \Big(d\mu_i^2 + \mu_i^2 (d\phi_i + g A^i)^2\Big)\,,
\ee
where $A^i$ are the four $U(1)$ gauge potentials, $X_i$ are exponentials
of the dilatonic scalar fields (defined in \cite{tenauthor})), and
the four direction cosines $\mu_i$ satisfy $\sum_i \mu_i^2=1$.
Writing these as
\be
\mu_1= \sin\tilde\theta\,,\quad \mu_2=\cos\tilde\theta\, \sin\tilde\varphi\,,
\quad \mu_3=\cos\tilde\theta\, \cos\tilde\varphi\, \sin\tilde\psi\,,\quad
\mu_4=\cos\tilde\theta\,\cos\tilde\varphi\, \cos\tilde\psi\,,
\ee
and substituting the scalar fields of our four-dimensional solution into the
expressions for the $X_i$ given in \cite{tenauthor}, we find
\bea
X_1 &=& \Big(\fft{H_1}{H_2}\Big)^{-3/4}\,,\qquad X_2=X_3=X_4=
   \Big(\fft{H_1}{H_2}\Big)^{1/4}\,,\nn\\
\tilde \Delta &=& H_1^{1/4}\, H_2^{3/4} \Big(H_1^{-1}\, \sin^2\tilde\theta +
  H_2^{-1}\, \cos^2\tilde\theta\Big)\,,
\eea
and the four
gauge potentials are given by $A_\1^1=A$ and $A_\1^2=A_\1^3=A_\1^4=0$.
The eleven-dimensional metric is therefore given by
\bea
d\hat s_{11}^2 &=& \tilde\Delta^{2/3}\,\, ds_4^2 +
  4 g^{-2} \tilde\Delta^{2/3}\,
\Big(\fft{H_1}{H_2}\Big)^{1/2}\, d\tilde\theta^2
  \nn\\
&&+4 g^{-2}\tilde\Delta^{-1/3}\, \Big[\Big(\fft{H_1}{H_2}\Big)^{3/4}\,
\sin^2\tilde\theta \, (d\phi_1+ \ft12 g A)^2 +
 \Big(\fft{H_1}{H_2}\Big)^{-1/4}\,
  \cos^2\tilde\theta \, d\Omega_5^2\Big]\,,
\eea
where
\be
d\Omega_5^2 = d\tilde\varphi^2 + \sin^2\tilde\varphi \,d\phi_2^2 +
  \cos^2\tilde\varphi\, (d\tilde\psi^2 + \sin^2\tilde\psi\, d\phi_3^2 +
     \cos^2\tilde\psi \, d\phi_4^2)
\ee
is the metric on a unit 5-sphere.  Here, $ds_4^2$ and $A$ are the metric and
gauge potential of the four-dimensional dyonic gauged black hole given in
(\ref{adsdyon}).

  Using the expression in \cite{tenauthor} for the 4-form field strength in
eleven dimensions, we obtain
\bea
\hat F_\4 &=& -g \Big[2 \Big(\fft{H_1}{H_2}\Big)^{1/2}\, \cos^2\tilde\theta +
 \Big(\fft{H_1}{H_2}\Big)^{-1/2} \,(1+2 \sin^2\tilde\theta)\Big]\,
\epsilon_\4 \nn\\
&& -2 g^{-1}\,\sin\tilde\theta\, \cos\tilde\theta\,
 (H_1^{-1}\, {*d}H_1
 -H_2^{-1}\, {*d}H_2)\wedge d\tilde\theta \nn\\
&&
 - 4 g^{-2} \Big(\fft{H_1}{H_2}\Big)^{3/2}\,
\sin\tilde\theta\, \cos\tilde\theta\, d\tilde\theta\wedge
  (d\phi_1+ \ft12 g A)\wedge {*F}\,,
\eea
where $\epsilon_\4$ is the volume-form of the four-dimensional metric and
$*$ denotes the Hodge dual in the four-dimensional metric.

  In the extremal limit, the metric $ds_4^2$ becomes AdS$_2\times S^2$,
which is the
base space for the fibre direction $d\phi_1 + \ft12 g A$.
For an appropriate Euclideanisation, the eleven-dimensional metric can
be viewed as a foliation of $T_{p,q}$ and $S^5$.

\section{Thermodynamics and the Wald Canonical Charge}

    Here we shall return to the intriguing
feature that we found when
studying the thermodynamics of the gauged dyonic black hole, namely that
the naively-expected first law $dM=TdS + \Phi_Q\, dQ + \Phi_P\, dP$
turned out not to hold except in the special cases enumerated
in (\ref{exception}).
In general, we found that we had to introduce a new pair of
thermodynamic conjugate variables $(X,Y)$ in the first law, as in
(\ref{modfirst}).  This is reminiscent of the situation for the
thermodynamics of black holes in conformal gravity \cite{Lu:2012xu,Lu:2013hx}.
However, in four-dimensional conformal
gravity, the black hole solution \cite{Riegert:1984zz} has an extra integration
constant on
account of the higher-derivative equations of motion, and hence it is
natural in that case to have to introduce the notion of ``massive spin-2 hair''
with its associated charge and potential \cite{Lu:2012xu}.  Our AdS dyon,
on the other hand,
has only three integration constants, which are related to the mass,
and the electric and magnetic charges.  Thus, the quantities $(X,Y)$
must be expressible as functions of $(M,P,Q)$.  This implies that it
would be possible to redefine the other thermodynamic quantities in such
a way that a ``standard''  first law, with no $X$ and $Y$, would satisfied.

This can be seen most easily in the case of the $\R^2$-symmetric  
solution (\ref{adsdyon2}),
whose thermodynamic quantities are given in (\ref{thermoquan}).
In this case, we see that $Y=Q/P$, and so if we redefine
\begin{equation}
\Phi_Q \longrightarrow \widetilde\Phi_q=\Phi_Q - \fft{X}{P}\,,\qquad
\Phi_P \longrightarrow \widetilde\Phi_P=\Phi_P + \fft{XQ}{P^2}\,,
\end{equation}
then the usual first law of thermodynamics would be recovered.  However, these
redefinitions are rather artificial, since the electric
and magnetic potentials are ostensibly well-defined quantities 
with no apparent {\it a priori} motivation for such a modification.  The
situation is even less clear in the case of the $S^2$-symmetric solutions 
(\ref{adsdyon}), since now $Y$ depends on $M$ as well as $Q$ and $P$, so the
necessary redefinitions would require seemingly unmotivated modifications 
to other quantities in addition to $\Phi_Q$ and $\Phi_P$.

A more satisfactory explanation would be if $(X,Y)$ admitted a physical
interpretation in their own right, as a scalar potential and charge.
At first sight this might seem problematic since, as can be seen from 
the asymptotic behaviour of the metric and
the scalar field, discussed in section 2,  the characteristic
behaviour for generic values of $(Q,P)$ appears to be the same as in 
either of the
special cases $P=0$ and $Q=0$.  And yet, the first law requires the
introduction of $(X,Y)$ in the generic case but not in
either of these special cases. 

  In order to settle this, we can
use the procedure
developed by Wald \cite{Wald:1993nt,Iyer:1994ys} to determine whether
there is some additional charge. Given the Lagrangian (\ref{lag}), one can
calculate the Noether charge associated with a Killing vector $\xi$. Following
\cite{Wald:1993nt,Iyer:1994ys}, the conserved charge takes the form
\bea
&&\cQ_{\xi}=\cQ^{\rm gravity}_{\xi}+\cQ^{\rm em}_{\xi},\nn\\
&& \cQ^{\rm gravity}_{\xi}=-\frac1{16\pi G}\, {*d\xi},\quad
\cQ^{\rm em}_{\xi}=-\frac1{16\pi G}(i_\xi A)e^{-\sqrt{3}\phi}\, {*F}\,,
\eea
where $i_\xi$ denotes the contraction of $\xi^\mu$ with a form field,
and so $i_\xi A= \xi^\mu A_\mu$.
In deriving this expression, we have chosen a gauge where
${\cal L}_{\xi}A_{\mu}=0$.
On the other hand, if we consider the variations of $g_{\mu\nu}$,
$A_{\mu}$ and $\phi$ on the solution space, we can define a closed
2-form whose differential, when integrated on the domain bounded by
infinity and horizon, leads to the first law of thermodynamics. As
explained in \cite{Wald:1993nt}, the integration of the closed 2-form at
infinity is identified as the infinitesimal Hamiltonian which generates
the Hamiltonian flow associated with the vector field $\xi$. The
infinitesimal Hamiltonian corresponding to the theory (\ref{lag}) is given as
\be
\delta {\cal H}=\int_{\infty}\Big(\delta \cQ_{\xi}-i_{\xi}\Theta\Big),
\quad i_{\xi}\Theta=i_{\xi}\Theta^{\rm gravity}+i_{\xi}\Theta^{\rm em}
  +i_{\xi}\Theta^{\rm scalar}\,,\label{deltaH}
\ee
where
\bea
i_{\xi}\Theta^{\rm gravity}&=&-\ft{1}{2}
\epsilon_{\alpha\beta\mu\nu}\xi^{\alpha}\Xi^{\beta}dx^{\mu}\wedge dx^{\nu}\,,
\quad
\Xi^{\beta}=\frac1{16\pi G}(g^{\mu\beta}g^{\alpha\nu}-
  g^{\mu\nu}g^{\alpha\beta})\nabla_{\alpha}\delta g_{\mu\nu}\,,\nn\\
i_{\xi}\Theta^{\rm em}&=&-\frac1{16\pi G}\Big(\delta(i_{\xi} A)
e^{-\sqrt{3}\phi}\, {*F}  +\Psi\delta F\Big)\,,\quad
d\Psi=e^{-\sqrt{3}\phi}\, i_{\xi}{*F}\,, \nn\\
i_{\xi}\Theta^{\rm scalar}&=&\frac1{16\pi G} \delta\phi\, i_{\xi}{*d}\phi\,.
\eea
Note that we are following the notation of Wald \cite{Wald:1993nt} here;
the symbol $\delta{\cal H}$ is not intended to imply that it is 
necessarily an integrable variation of a Hamiltonian ${\cal H}$.

  It should be noted also that we do not vary the gauge coupling parameter
$g$ here. In the gauge ${\cal L}_{\xi}A_{\mu}$=0, it can be shown that
$d(i_{\xi} A)$ is gauge invariant. Therefore $i_{\xi} A$ can only be
shifted by a constant under the gauge transformation. We fix this residual
gauge degree of freedom by requiring $i_{\xi} A|_{\infty}=0$.

For our spherically-symmetric dyon, from the large $r$ expansion discussed
in section 2, we have
\begin{equation}
\tilde{h}-h=\Delta_1+\frac{\Delta_2}{\rho}+\cdots,\qquad
\phi=\frac{\phi_1}{\rho}+ \frac{\phi_2}{\rho^2}+\cdots\,,\label{asymbehave}
\end{equation}
where $\rho$ is defined by (\ref{schcoord}) and
\begin{eqnarray}
&&\Delta_1=\ft14g^2\phi_1^2\,,\qquad
\Delta_2=\ft23g^2\phi_1\phi_2\,,\cr
&&\phi_1=-\frac{2\sqrt{3}\mu(\beta_1-\beta_2)(1-\beta_1\beta_2)}{
\gamma_1\gamma_2},\quad \phi_2=\frac{2\sqrt{3}\mu^2(\beta_1-\beta_2)
  \gamma}{\gamma_1^2\gamma_2^2}\,.
\end{eqnarray}
Taking $\xi$ to be the time-like Killing vector
$\xi=\partial/\partial t$, we find
\be
\delta{\cal H}=\int_{\infty}\Big(\delta \cQ_\xi-i_{\xi}\Theta\Big)=
\delta M + Z\,,
\ee
where $M$ is the mass of the dyon given in (\ref{mass}) and $Z$ is
given by
\begin{equation}
Z=-\ft12\delta\Delta_2+\ft12g^2\phi_2\delta\phi_1+\ft14g^2\phi_1\delta\phi_2=
\ft1{12} g^2 (2\phi_2 \delta\phi_1 - \phi_1 \delta\phi_2)\,.
\end{equation}
Here we have set Newton's constant $G=1$. It is clear that $Z$
vanishes if $\phi_2=c \phi_1^2$, for any numerical constant $c$.
Thus when $\beta_1=\beta_2$ and hence $\phi_i=0$, or when
$\beta_1=0$ or $\beta_2=0$, for which $\phi_2=\mp\phi_1^2/(2\sqrt3)$,
the quantity $Z$ vanishes and the usual thermodynamics holds.
For generic $\beta_i$, we find
\begin{equation}
Z=- X dY\,.
\end{equation}
where $(X,Y)$ are precisely the quantities introduced in section 2 in
order to make the first law of thermodynamics to work.  The procedure
above shows that the quantities $(X,Y)$ depend only on the
asymptotic parameters and are independent of properties at the
horizon.

   The inclusion of a scalar charge in the thermodynamics
was discussed in the context of asymptotically-flat black holes in
\cite{Gibbons:1996af}.  The details were rather different there, with the
scalar charge being a modulus corresponding to the value $\phi_\infty$ 
of the scalar
field at infinity, on
account of the different asymptotic behaviour in that case.  
In contrast, our solution always has $\phi_\infty=0$, the
stationary point of the scalar potential.
A phenomenon that is more similar to the one we have encountered
in this paper arose for asymptotically-AdS black holes in conformal gravity
\cite{Lu:2012xu,Lu:2013hx}, where the infinitesimal Hamiltonian is not
integrable and the term $\int_{\infty}i_{\xi}\Theta$ leads to the
inclusion of extra canonical pairs.  In fact, the parallel to
\cite{Lu:2013hx} is stronger, since in \cite{Lu:2013hx} extra canonical
pairs beyond the total number of integration constants had to be
introduced also.

In general the scalar $\phi$ approaches the
asymptotic AdS as in (\ref{asymbehave}), where $\phi_1$ and
$\phi_2$ can in principle depend on other coordinates as well, but are
constants for our static dyon.  There are three types of boundary
conditions that preserve all the asymptotic AdS symmetries. The first
two are either $\phi_1=0$ or $\phi_2=0$
\cite{Breitenlohner:1982bm,Klebanov:1999tb}.  The third is where
$\phi_2=c\phi_1^2$ \cite{Hertog:2004dr} (see also \cite{Hertog:2004ns}.)
Interestingly, in all these three cases the quantity $Z$ vanishes.
Thus, the quantity $Z$ we obtained via the Wald procedure describes
scalar hair that breaks some of the asymptotic AdS symmetries.

\section{Conclusions}

In this paper, we constructed the static AdS dyonic black hole in
four-dimensional maximal gauged supergravity, where one of the $U(1)$
gauge fields carries both electric and magnetic charges. If the
gauge coupling constant is sent to zero, the solution reduces
to the well-known asymptotically-flat Kaluza-Klein dyon.  Our solution
can be lifted to eleven dimensions, where it provides a new solution of
M-theory.  Since the four-dimensional solution is asymptotic to AdS,
the gauged dyon provides a new background for exploring the
AdS/CFT correspondence.

 An intriguing feature arises when we study the thermodynamics of the
gauged dyonic black hole. The naively-expected first law $dM=TdS + \Phi_Q dQ
+ \Phi_P dP$
does not hold except in the special cases enumerated in (\ref{exception}).
In general, we find that we have to introduce a new pair of
thermodynamic conjugate variables $(X,Y)$ in the first law, as in
(\ref{modfirst}).  By examining the asymptotic fall-offs and
employing the Wald procedure for calculating the conserved quantities
associated with the time-like Killing vector, we find that $X$ and $Y$
describe the potential and charge of the scalar hair that breaks some of
the asymptotic AdS symmetries.  This is the first explicit analytical
example of a
charged black hole with  neutral scalar charge in gauged
supergravity.  It may be that a more general four-parameter class of 
static gauged dyonic black holes exists, where the scalar charge $Y$
can be specified independently of $M$, $Q$ and $P$, with our solution 
being a special case where $Y$ is a specific function of $(M,Q,P)$.
If such a more general solution were known, it could shed further light
on the physical meaning of the quantities $X$ and $Y$, and the 
possible interpretation of including an $XY$ term in a Legendre transformation
of the free energy.  Since our AdS dyon is a solution
to maximum gauged supergravity and is dual to an ${\cal N}=8$, $D=4$,
superconformal field theory, it would be of great interest to study the gauge
duals of the scalar hair.

\section*{Acknowledgement}

We are grateful to Gary Gibbons and Andy Strominger for useful discussions.
H.L.~is supported in part by the NSFC grants 11175269 and 11235003.
C.N.P.~is supported in part by DOE grant {DE}-{SC}0010813.

\end{document}